\documentclass[sigconf]{acmart}

\AtBeginDocument{%
  \providecommand\BibTeX{{%
    \normalfont B\kern-0.5em{\scshape i\kern-0.25em b}\kern-0.8em\TeX}}}

\usepackage[utf8]{inputenc}
\usepackage{url}
\usepackage{todonotes}
\usepackage{amsmath}
\usepackage{amssymb}
\usepackage{mathtools}
\DeclarePairedDelimiter{\ceil}{\lceil}{\rceil}

\setcopyright{acmcopyright}
\copyrightyear{2020}
\acmYear{2020}

\acmConference[IWMUSC'20]{1st SIGSPATIAL International Workshop on Modeling and Understanding the Spread of COVID-19}{November 03, 2018}{Seattle, Washington}

\begin{document}
  
\title{Infection Risk Score: Identifying the risk of infection propagation based on human contact}
\author{Rachit Agarwal}
\email{rachitag@iitk.ac.in}
\affiliation{IIT Kanpur, India
}
\author{Abhik Banerjee}
\email{abanerjee@swin.edu.au}
\affiliation{Swinburne University, Australia}

\renewcommand{\shortauthors}{Agarwal and Banerjee}

\begin{abstract} 
A wide range of approaches have been applied to manage the spread of global pandemic events such as COVID-19, which have met with varying degrees of success. Given the large-scale social and economic impact coupled with the increasing time span of the pandemic, it is important to not only manage the spread of the disease but also put extra efforts on measures that expedite resumption of social and economic life. It is therefore important to identify situations that carry high risk, and act early whenever such situations are identified. While a large number of mobile applications have been developed, they are aimed at obtaining information that can be used for contact tracing, but not at estimating the risk of social situations. In this paper, we introduce an infection risk score that provides an estimate of the infection risk arising from human contacts. Using a real-world human contact dataset, we show that the proposed risk score can provide a realistic estimate of the level of risk in the population. We also describe how the proposed infection risk score can be implemented on smartphones. Finally, we identify representative use cases that can leverage the risk score to minimize infection propagation.
\end{abstract}
\keywords{
Infection risk score, Contact Tracing, Mobile Computing, Internet of Things, Mobile Health
}

\maketitle

\section{Introduction}

21st century has already been witness to multiple pandemics in the first two decades, with the biggest being COVID-19 caused by the SARS-CoV-2 virus. The unprecedented spread of the COVID-19 has led to global efforts by governments to contain the pandemic and to limit the impact of the virus on human society. As with any other infectious disease, the efforts to contain the virus largely focus on \textit{(i)} minimizing human-to-human contact by enforcing people to maintain a certain distance with others (also known as \textit{\textbf{social distancing}}),  \textit{(ii)} minimizing economic activity (also known as \textit{\textbf{lockdown}}) for a certain period in geographical regions, such as cities, states or even entire countries, and \textit{(iii)} by performing \textit{\textbf{contact tracing}}, which involves tracking the disease spread by identifying the contacts of the confirmed cases. However, these efforts have been met with varying degrees of success, and the authorities have been trying to use technology as much as possible to elevate their efforts~\cite{khazan_2020}.

With the rise of the Internet of Things (IoT) and mobile health~\cite{Wood2019} (also referred to as \textit{mHealth}), there has been a growth in the number of possibilities related to not only understanding the environment but also detecting  diseases early. With regards to the COVID-19 pandemic  in particular, governments around the world have looked to leverage the use of smartphone applications for limiting the spread of the disease, given the ubiquity of smartphone usage. While many of these applications focus on providing up-to-date information about the spread of the disease, other applications aim to notify users in real-time when they come in contact with an infected person~\cite{Islam2020}. These infection tracking applications use a variety of sensors embedded in a smartphone to help detect the transmission in real-time. A common type of sensor used is Bluetooth Low Energy (BLE), which can be used for proximity detection. Multiple applications that leverage BLE for the purpose of monitoring the growth of the COVID-19 pandemic have been introduced in various countries. In India, the Aarogya-Setu Application~\cite{AarogyaSetu} informs how many infected people are within a certain distance of a person using the application by matching with national database of the infected people. In Australia, the COVIDSafe application~\cite{covidsafeapp2020} provide notifications to the users if their contact is detected with a confirmed infected person. Similar applications have been developed by the governments of many other countries. Further, a collaboration between Apple Inc. and Google has led to the development of an Exposure API that enables developers to build various applications using which application users can know if they came into contact with other infected people~\cite{ExposureNS}. Apart from smartphone applications, other types of technologies are also used to help in the cause of containing pandemic. These include the use of SwipeSense technology to track use of medical equipment and to track whether hospital staff wash their hands regularly\footnote{https://www.cnbc.com/2020/08/02/hospitals-tracking-covid-19-with-badge-sensors-swipesense-technology.html}.

Despite the technological innovations and advancements, the use of applications, such as those described above, and technologies for managing and controlling the spread of the infection is challenging due to multiple reasons. Firstly, a person carrying the disease may not show any symptoms for a long period (e.g. when the infection is in the incubation period). As a result, any close contacts with other people would not be detected as being risky by infection tracking applications, and hence would not help in containing the spread of the infection. Indeed, in the case of COVID-19, a significant fraction of cases have been identified as asymptomatic for the entire duration of infection~\cite{asymptomatic_covid}. These also contribute to community spread of the disease, which can lead to exponential growth in the number of infections. Secondly, once someone is confirmed as infected, he/she is typically isolated and is not allowed to get involved in any social activities until fully recovered. Thirdly, existing methods for managing infection spread are primarily reactive. Counter-measures are often taken after a person is confirmed to be infectious. Subsequently, authorities proceed with counter measures such as lockdown of the specific geographical region. Thus, currently, the scope for detection of the infection spread and its management is limited. Therefore, there is a need for an early estimate of the potential risk in a geographical region to enable  authorities to act quickly.

For risk estimation to be effective, it needs to identify people who have greater exposure to the infection, quarantine the exposed people, identify regions with potentially high exposures, and declare a region as hot-spot even before the outbreak happens in that region. Additionally, the risk estimation measure should also be able to identify situations which are likely to lead to transmissions even when there are not any confirmed presence of the known infections. Finally, any such risk estimation needs to be adaptable to a wide range of technology platforms. While a notion of risk score has been introduced as part of the Exposure API by Apple Inc and Google, its main drawbacks is that it only provides a risk measure based on confirmed exposures to infections. 



In this paper, we present a risk score that can be used to assess the risk for individuals based on their contact events. A key novelty of the proposed risk score is that it estimates the risk propagation, unlike existing literature that only assess immediate risk. The proposed risk score can be used to assess the level of risk within geographical regions, enabling authorities to act early to contain a potential outbreak. Further, monitoring the risk score can also help individuals take actions. 
In particular, the key contributions of this paper are as follows: 
\begin{itemize}
    \item \textit{\textbf{Infection risk score}}: We introduce a risk score that estimates infection propagation by monitoring contact events among individuals.The risk score takes into consideration factors such as the contact proximity, transmission likelihood and vulnerability to a disease. 
    \item \textit{\textbf{Evaluation using realistic dataset}}: We evaluate the infection risk score using a real-world human contact dataset that has previously been using to study infection propagation. 
    Our results show that potentially risky situations are well captured using the infection risk score.  
    \item \textit{\textbf{Adaption of risk score using smartphones}}: We provide detailed description on how smartphones can be used to implement the infection risk score to track infections. 
\end{itemize}
Finally, we also discuss how the accuracy of infection risk score can be improved by incorporating contextual information, and also present a discussion on potential use cases of the risk score for managing infection spread.

The rest of the paper is organised as follows. Section~\ref{sec:rw} provides detailed survey of related techniques and existing metrics used to quantify risk and exposure. Section~\ref{sec:model} provides details of the proposed risk model. In Section~\ref{sec:ev} we evaluated the model using real data. Section~\ref{sec:middleware} provides the details on the propose version of the smartphone application. This is followed by perspective uses of the risk score in section~\ref{sec:fd}. We finally conclude in Section~\ref{sec:conc}.

\section{Related work}\label{sec:rw}

Recent studies focusing on containing pandemic can be mainly classified into three broad groups: survey based studies, IoT based studies and epidemic model based studies. 

In survey based studies, in \cite{mhango2020}, authors report that factors such as contact with infected person, work overload, medical history of the person, and if the person wore Personal Protective Equipment (PPE) or not play an important role in determining the risk of infection transmission of COVID-19. Similarly, to identify potential exposure, WHO uses risk assessment forms to determine the risk of exposure. Here they ask questions related to if the person wore the PPE as recommended or not~\cite{who2020CovidHealth}. 

In IoT based studies, there is increased focus on smartphone based infection detection. 
Many applications and IoT Devices are available that perform contact tracing using proximity checks. A survey of some of these application is present in~\cite{Islam2020}. We do not survey these applications again and instead present, in brief, new applications and devices that have come-up since the publication of~\cite{Islam2020}. 
Recent applications and devices includes 
EasyBand, a wearable device that vibrates when a marked (infected) Easyband comes in close proximity~\cite{shukla2020privacy}. Nonetheless, it has issues related to centralized control and communication. In~\cite{Jeong2019}, authors used magnetometer based proximity detection, while in~\cite{nguyen2020epidemic} authors used multiple sensors to improve the distance estimation accuracy. Such techniques fail in the case when a smartphone lacks certain required sensor. Further, these applications achieve \textit{privacy by architecture} and not \textit{privacy by design}. Many recent application and IoT devices claim to follow privacy guidelines such as those mentioned in~\cite{shukla2020privacy}. These applications and devices include: \textit{(i)} Pan European Privacy-Preserving Proximity Tracing (PEPP-Pt)\footnote{https://www.pepp-pt.org} that uses anonymized ID for communication, \textit{(ii)} TraceSecure that uses secret sharing technique to identify proximity~\cite{bell2020tracesecure}, and \textit{(iii)} proximity-based privacy-preserving contact tracing (P$^3$CT) that uses ambient signature protocol~\cite{ng2020epidemic}. Again, while these applications are privacy preserving, they achieve privacy by architecture. In summary, these studies model risk using factors such as distance~\cite{AarogyaSetu,shukla2020privacy,Jeong2019,nguyen2020epidemic,bell2020tracesecure} and duration~\cite{ExposureNS}. These works mainly use either BLE or magnetometer to estimate distance from neighbor. Nonetheless, these works do not quantify the risk, and instead, just provide an estimate of whether a person was in contact with some other person or not.

On the other hand, from epidemic modeling point of view, there are many studies that quantify risk using different parameters such as: size of cough droplets, rate of cough, volume of particles generated, concentration of pathogens, max distance covered by pathogen in air, pathogen particles lost due to temperature and humidity, time an infected person stayed at a given location, duration of contact with susceptible person, and his pulmonary rate~\cite{Shahzamal2017}. Most of these factors, until now, cannot be estimated using smartphone. Instead, disease specific average values for these factors can be used as constants while modeling risk score. In~\cite{Shahzamal2017}, authors estimated risk as an aggregation of risk score for both when a person comes in direct contact with other person and when a person gets infected indirectly (a case of \textbf{\textit{community spreading}}). 

\section{Infection Risk Score}\label{sec:model}
In this section, we present the \textit{infection risk score} that quantifies risk of catching an infection. Our score considers exposure to pathogen and context in a social network setting. For convenience, \textit{infection risk score} is referred by the term \textit{risk score} in the remainder of the paper.  

\subsection{Network model: Modeling the population as a temporal network}
For any geographical area, we consider the population to be represented by a temporal graph $G$ such that $G(V_t,E_t)$ is a temporal snapshot at time $t$ that is created by individuals in a given area $A_a$. For the purpose of this paper, we consider that the risk score computation for individuals is done using mobile apps, and hence, each individual is represented using smartphones. Here $V_t$ is the set of smartphones communicating and active at time $t$ and $E_t$ is the set of edges that exists between smartphones in $V_t$. Let $\Delta t$ be the time difference between two consecutive temporal snapshots of $G$. For our model we assume that if two people are in contact, for say 10 epochs, then the edge between them is persistent over $\frac{10}{\Delta t}$ snapshots of the graphs. 
Each person $i\in V_t$ has a location, $l_t$, marked by latitude and longitude pair such that $l_t=(la_{i,t}$, $lo_{i,t})$. Given the interactions, at time $t$, each person $i$ has a neighborhood, $N_{i,t}$ where each person $j\in N_{i,t}$ has an edge (in $E_t$) to the person $i$ and is $d_{i,j,t}$ distance apart. Here $d_{i,j,t}<\theta_d$ i.e., $i$ and $j$ are within communication range and at maximum $\theta_d$ distance apart.

\subsection{Risk score parameters}
In this section, we identify the key factors that impact infection propagation. 

\begin{enumerate}
    \item \textbf{Exposure caused by a neighbor}: Communicable diseases such as COVID-19 generally spread when a person $i\in V_t$ comes in close proximity with a infected person (person $j$) or touches the surface that infected person has touched~\cite{Shahzamal2017}. In such a case, the person $i$ is exposed to pathogens from the infected person, which can lead to infection spread. The exposure to a neighboring individual is an key factor determining the likelihood of a transmission event from a neighbor, and we term this as the \textbf{neighbor exposure}. For the scope of the current paper, we limit our discussion to the the exposure caused when an infected person come in close proximity, although this may easily be extended to include other modes of propagation. 
    
    To determine how the neighbor exposure impacts the spread of infection, we consider that an infected neighbor $j$ exhales $n_{i,j,t}\in \mathbb{R^+}$ pathogens and these pathogens are homogeneously distributed within the permissible $\theta_d$ distance. Further, we consider the following assumptions: \textit{(i)} there is no loss in pathogens, \textit{(ii)} each time same number of pathogens are exhaled, and \textit{(iii)} between two consecutive temporal snapshots of the graph (i.e., $G(V_t,E_t)$ and $G(V_{t-\Delta t},E_{t-\Delta t})$), a person $i$ stays in contact with person $j$ for the $\Delta t$ time. In such a scenario, the 
    exposure to an infectious disease of the person $i$ at time $t$ with respect to a particular neighbor $j$ is given by $E_{i,j,t}=\Delta t\times n_{i,j,t}$. 
    
    In ideal conditions, if a neighbor $j$ is not infected, i.e., he/she does not cough, and wears proper protective gears such as face mask or face shields, $E_{i,j,t}=0$ because there are no pathogens exhaled by $j$. In such a case, the whole idea of maintaining social distancing even when people are not infected would fail and susceptible people would be deemed harmless. On the other hand, if some neighbor is infected and coughing badly, $E_{i,j,t}>>0$. In this case, other people would ideally limit from meeting the infected person. In such a situation also, barring the infected person, other susceptible people would continue their physical social activities. Let $r_{j,t-\Delta t}$ be the risk score of the neighbor at time $t-\Delta t$. To account for above mentioned aspects and ensure that social distancing is enforced between susceptible people also, we add the previous instance risk score of the neighbor to the exposure caused due to the neighbor, i.e., $E_{i,j,t}=\Delta t\times n_{i,j,t} + r_{j,t-\Delta t}$.
    
    \item \textbf{Neighbor weight}: We define the \textbf{neighbor weight} as the likelihood that an individual in the vicinity is infectious. Since we aim to estimate the risk even in situations where confirmed infections are not known, the neighbor weight can be estimated based on multiple contextual parameters. For instance, in the case of communicable diseases such as COVID-19, if a neighbor is from a hot-spot area or has a history of the disease then the risk of getting infection from the neighbor is high because the neighbor is coming from a containment zone. Further, impact of diseases like COVID-19 is high on people who have a weak immunity either due to age or have chronic diseases like kidney failure and diabetes. On top, if a person is staying indoor, with a poor ventilation chances of spreading the disease and getting infected increases manyfold~\cite{Smieszek2019, Vuorinen2020}. In~\cite{Smieszek2019}, authors recommend that proper ventilation indoor can reduce infections up-to 60\%. Nonetheless, for COVID-19, different countries have different statistics, for example, India having relatively younger population, middle age people are more infected while more older people have died. Let $w_{j,t}$ be the \textbf{\textit{weight}} such that $w_{j,t}\in[0,1]$ that identifies such contextual information of the neighbor. Summed over all the neighbors of the person $i$ at time $t$, the total exposure of $i$ from its neighbors $j\in N_{i,t}$is thus given by equation~\ref{eq:totalExposureContext}. 
    \begin{equation}\label{eq:totalExposureContext}
        E_{i,t}=\sum_{j\in N_{i,t}} w_{j,t}\times (E_{i,j,t}+r_{j,t-\Delta t})
    \end{equation}
\end{enumerate}

\subsection{Risk score formulation}
In addition to the neighbor weight and exposure, we define \textbf{vulnerability} as the likelihood that an individual exposed to risky situations continues to be at risk. 
At any time $t$ the risk score of an individual $i$ is the dependent on the risk score at $t-\Delta t$, his vulnerability 
, and the  exposure from the neighbors at $t$. 
The total risk, thus, is given by equation~\ref{eq:total}.

\begin{equation}\label{eq:total}
    r_{i,t}= \frac{v_{i,t}\times r_{i,t-\Delta t} + \sum_{j\in N_{i,t}} w_{j,t}\times (E_{i,j,t}+r_{j,t-\Delta t})}{1+ \sum_{j\in N_{i,t}} w_{j,t}}
\end{equation}

Here the denominator is the normalization factor. A disease usually has a period between when the person $i$ gets infected from the disease and time when he becomes an active spreader of the disease. For example, for COVID-19, the median incubation period is around 5 to 6 days\footnote{\url{https://www.mohfw.gov.in/pdf/DGSOrder04of2020.pdf}}. In our scenario, even if a person comes in contact with a person for whom the disease is still in incubation period, the risk exposure is equally high as compared to meeting a person who is an active spreader. Thus, our model does not consider the incubation period.  

From the equation~(\ref{eq:total}), the value of $r_{i,t}\in\mathbb{R^+}$. If the person is taken into isolation (i.e., no interaction with neighbors) after getting infected, his risk score will decrease with a factor $v_{i,t}$ and will eventually decay and reach minimum in $t=\ceil{\frac{r_{i,t-\Delta t}}{v_{i,t}}}$ time instances. This accounts for the fact that risk to and from such people is minimized when they are in isolation. For simplicity, at $t=0$ (or the initial condition) for all people we assign them as susceptible and their risk score to $r_{i,0}=1$. As the actual infection state of a person is unknown, the idea of social distancing mandates to maintain a certain distance even if the person is susceptible. Maintaining social distancing reduces the possibility of getting infected. We assign a non zero value to $r_{i,0}$ to ensure that social distance is maintained and our model captures it. For simplicity, let $r_{i,0}=1$. As and when a person is officially tagged infected, we assign $r_{i,t}=2$. 
Note that, a low value of $r_{i,t}$, is achieved when all the neighbors are susceptible. For a new person joining in, we assume that he is a susceptible person. 

Our method only considers ego network of a person for the calculation of the risk score. This enables all the smartphones involved to compute their individual risk scores simultaneously. 

\section{Evaluation and Validation}\label{sec:ev}

In this section we provide an evaluation and validation of our risk model using a real-world dataset. 


\subsection{Dataset}
While there are many datasets which have previously been used to study epidemic spread, specially smartphone based datasets that use Call Detail Records (CDRs) and GPS location information, ~\cite{Blondel2015}, they are \textit{(i)} not widely used~\cite{Oliver2020}, and \textit{(ii)} mostly generated from a random population sample which do not reflect true neighborhood size. Instead, we use a dataset of 789 individuals (including students and teachers) obtained on a single day in an American high school that has 158 rooms~\cite{salathe2010high}, which has previously been used to study spread of infectious diseases~\cite{Smieszek2019}. Here each point of interest (POI) is considered to be a room in the school. The dataset is mainly used to study human contact network for infectious disease transmission. The dataset is collected between 6AM to 4:30PM at an interval of 20 seconds. The granularity of positioning information available is at the level of rooms, and hence, each individual is geo-tagged with the room ID they are in at a particular epoch. We consider that contact events occur between individuals whenever they are in the same room, and all individuals present in a particular room at a given epoch are connected to each other. 

\begin{figure}[ht]
    \includegraphics[width=\columnwidth]{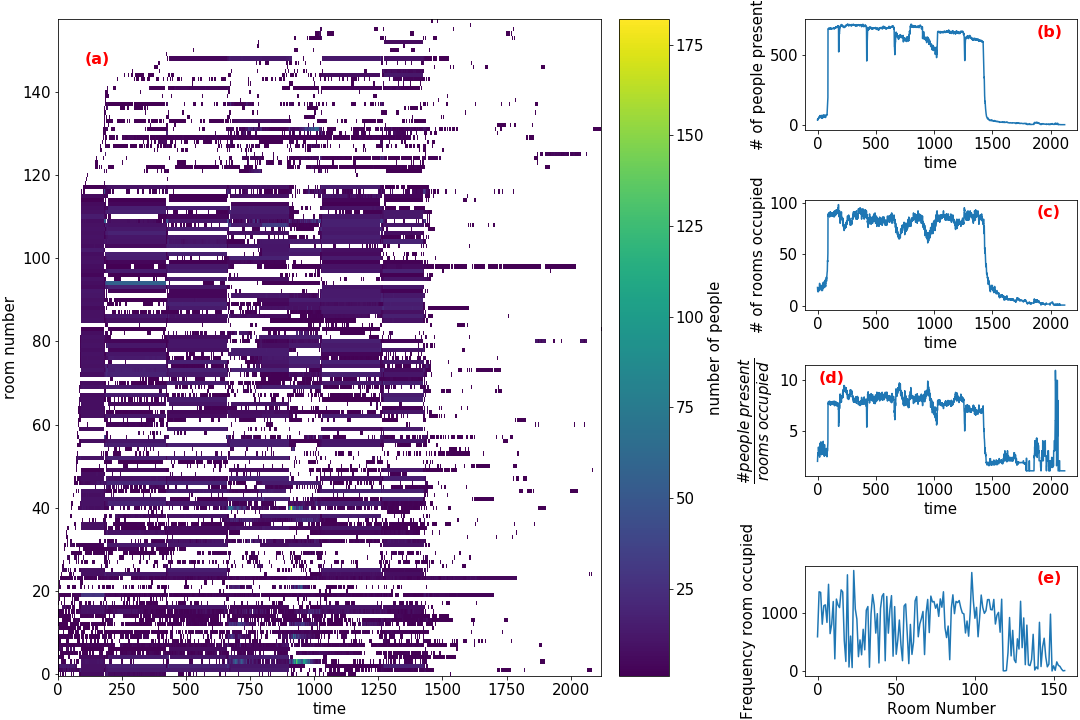}
    \caption{Temporal distribution of people in the rooms. (a) Heatmap showing number of people in each room over time. The white color represents empty room at the particular epoch. (b) Total number of people present in the school over time. (c) Total number of room occupied in the school over time. (d) average density of each room. (e) Number of times a room is occupied.}
    \label{fig:degreeEachRoom}
\end{figure}

The temporal distribution of individuals in the dataset is shown in Fig.~\ref{fig:degreeEachRoom}. Fig.~\ref{fig:degreeEachRoom}(a) shows the heatmap of number of people present in a room at different epochs. 
The white color represents that nobody was present in a room at the particular epoch. Fig.~\ref{fig:degreeEachRoom}(b) presents total number of people in a school at a given epoch. A sudden increase and a sudden drop in the number of people accounts for the beginning of the school in the morning when people arrive, and the end of the day, when they went back from school. Fig.~\ref{fig:degreeEachRoom}(c) presents the maximum number of rooms occupied by people. Note that at maximum only $\approx$62\% rooms are occupied. Fig.~\ref{fig:degreeEachRoom}(d) presents ratio between number of people in the school and rooms occupied at a given time. The maximum average density of people in a room is 9. A sudden increase at the end of the day is because most of the people were present in a single room. Fig.~\ref{fig:degreeEachRoom}(e) presents number of times a given room was occupied during the data collection period. From the figure we infer that (i) some rooms were always empty and nobody went to those rooms, (ii) the entire population is concentrated in only a few rooms and after certain time period there is an exponential decrease in the population indicating the end of classes in the school, (iii) during the day, rooms gradually start to fill up and there is an exponential rise in the population size.

\subsection{Dynamics of epidemic spread on the contact network}


We evaluate the proposed risk score using both SI (Susceptible-Infected) and SIS (Susceptible-Infected-Susceptible) models. 

Let $S_{i,t}$ be the fraction of people that are susceptible in a region $i$ at time $t$, $I_{i,t}$ be the fraction of population that is infected in a region $i$ at time $t$, $N_{i,t}=1$ be the faction of total population in the region at time $t$, $\beta_i$ be the infection rate in the region $i$, and $\gamma_i$ be the recovery rate in the region $i$. Note that at any given point of time $N_{i,t}=S_{i,t}+I_{i,t}$ because we consider only two states, susceptible and infected. The change in the fraction of susceptible and infected people over time is given by equation~(\ref{eq:epid})~\cite{Hethcote2000}. Here the underlying assumptions are that there is a homogeneous mixing of the population and no birth and death happens (the total population is fixed).

\begin{equation}\label{eq:epid}
    \begin{matrix}
        \frac{dS_{i,t}}{dt} =& -\frac{\beta S_{i,t}I_{i,t}}{N_{i,t}} + \gamma I_{i,t} \\ 
        \frac{dI_{i,t}}{dt} = & - \frac{dS_{i,t}}{dt}
    \end{matrix}
\end{equation}

\subsection{Results}

Currently, the exact behavior of exposure and vulnerability parameters for pandemics such as COVID-19 is not known. Further, as the dataset is POI based, actual distances are also not available in the dataset. Thus, we assume that the exposure parameter for each person is normally distributed with $\mu=0.5$ and $\sigma=0.1$. Further, the vulnerability parameter is also normally distributed with $\mu=0.5$ and $\sigma=0.2$. 

\begin{figure*}
    \includegraphics[width=0.95\textwidth]{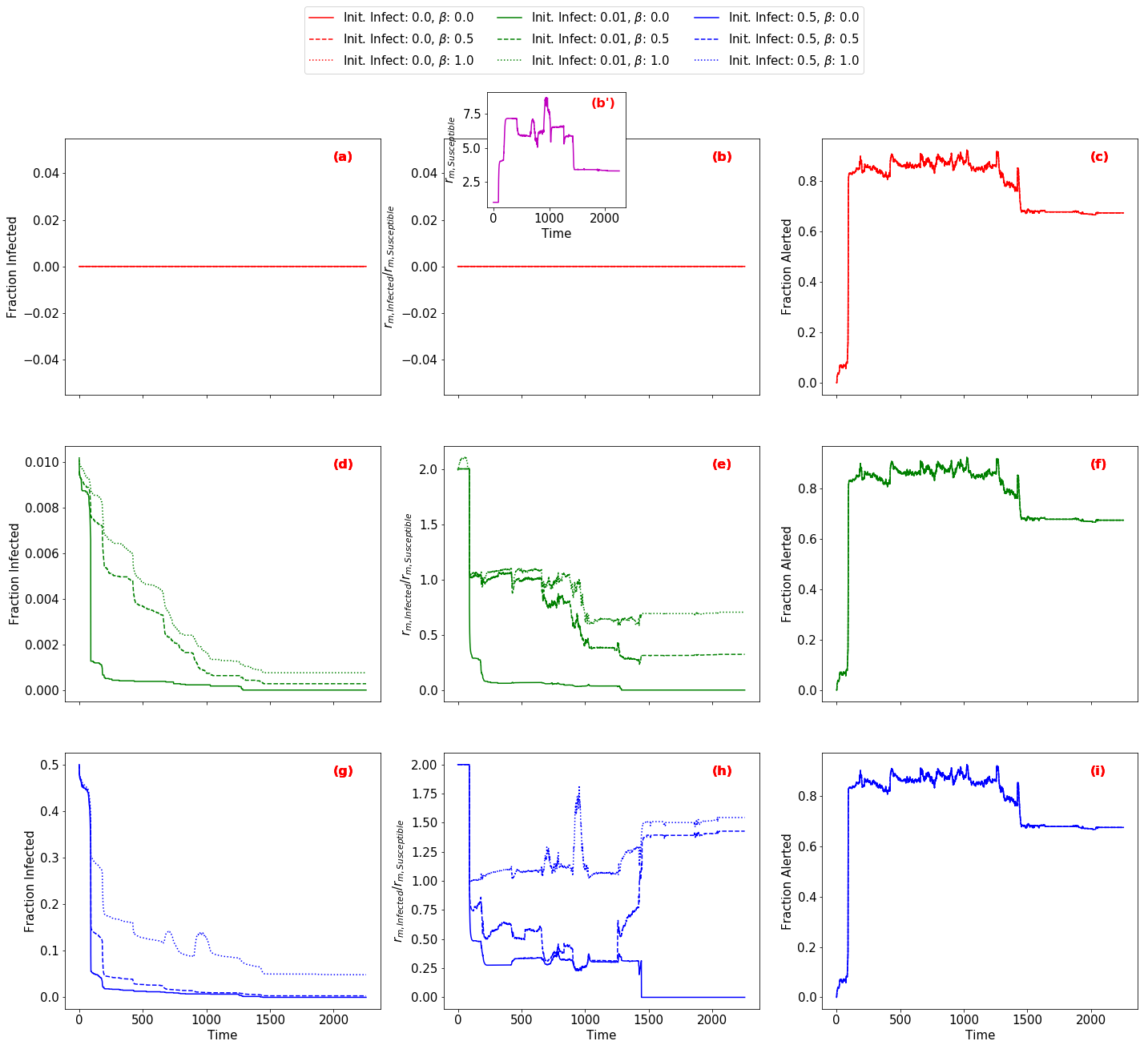}
    \caption{Results for SIS epidemic model, $\beta\in\{0.0,0.5, 1.0\}$, $\gamma_i=0.75$ and $I_{i,0}\in\{0.0, 0.01, 0.5\}$}
    \label{fig:SISGaussian}
\end{figure*}

For our analysis we study following three aspects using different initial condition, infection rate, and recovery rate. First, we identify fraction of people who are identified infected using the SIS and SI epidemic models. This helps us understand the infection spread over time in the population and understand the \textit{dynamics on} the contact network. Second, we measure the ratio between the median risk scores of infected people and susceptible people. A ratio more than one indicates that the risk score of infected people is more, as intended. A higher ratio implies that the risk score can be used to better identify 
people who are exposed to infection and have high probability to get infected. 
When there are no infections in a neighborhood, this value tends to 0. Third, we study the fraction of people that are alerted using our model. 

To test and study the above-mentioned aspects, as an initial condition, the values for $I_{i,0}$, $\beta_i$ and $\gamma_i$ used are $I_{i,0}\in\{0.0, 0.01, 0.5\}$, $\beta_i\in\{0.0, 0.5, 1.0\}$ and $\gamma_i\in\{0.0, 0.75\}$. $I_{i,0}=0.0$ states that there are no initial infections in the region. $\beta_i=0.0$ states there are no transmission happening and the disease does not spread via contact. On the other hand, $\beta_i=1.0$ would state that the disease is highly contagious. Similarly, $\gamma_i=0.0$ would state that there is no recovery which is equivalent to SI type epidemic model. $\gamma_i=0.75$ would mean that the recovery rate is 75\% (i.e. similar to 
the recovery rate of COVID-19 patients in India\footnote{https://www.financialexpress.com/lifestyle/health/indias-covid-19-recovery-rate-nears-75-case-fatality-rate-one-of-the-lowest-globally-at-1-86/2063108/}). The results presented here are averaged over 50 simulations runs and conducted using python.

\begin{figure*}
    \includegraphics[width=0.95\textwidth]{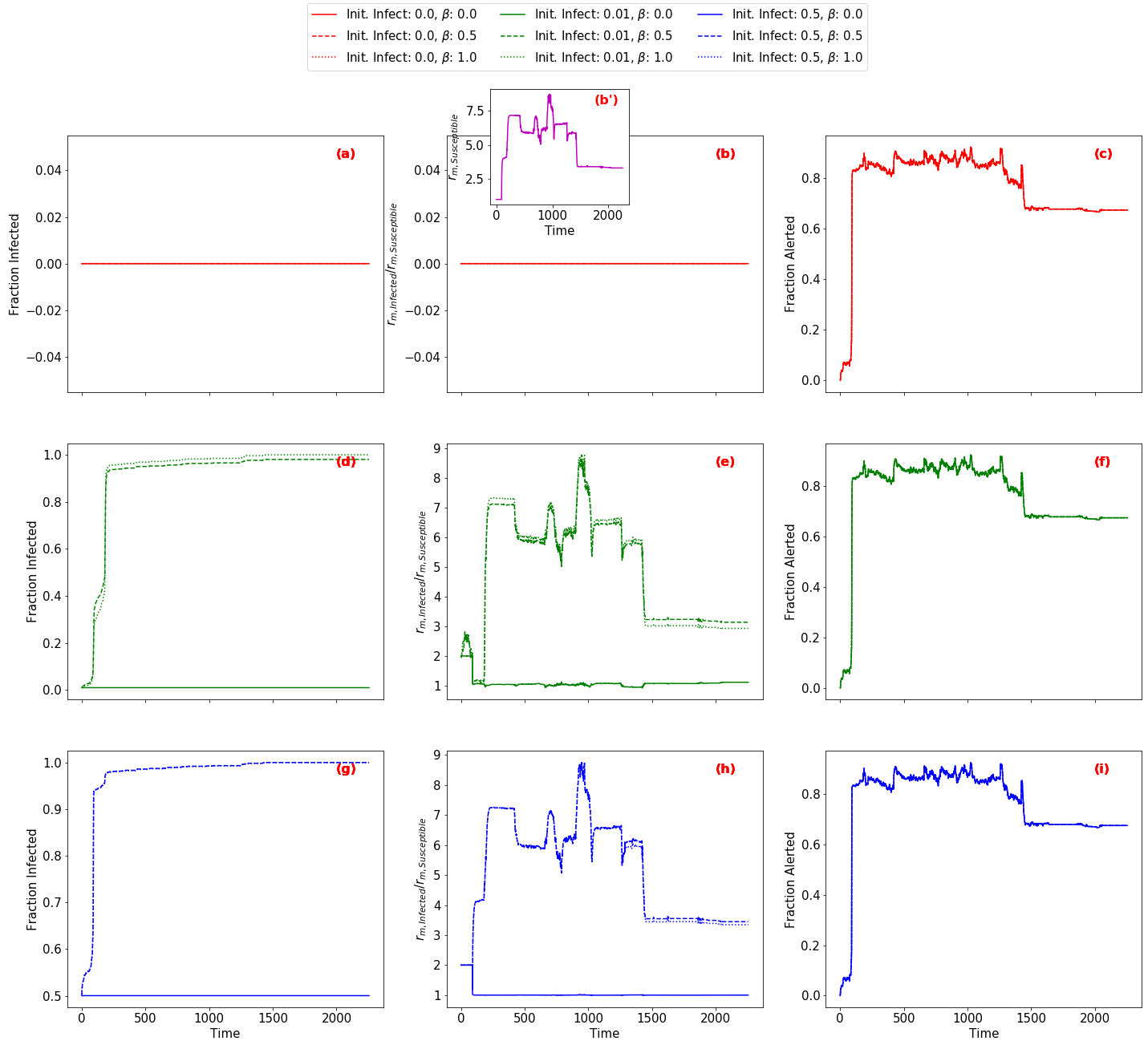}
    \caption{Results for SI epidemic model, $\beta\in\{0.0,0.5, 1.0\}$, $\gamma_i=0.0$ and $I_{i,0}\in\{0.0, 0.01, 0.5\}$}
    \label{fig:SIGaussian}
\end{figure*}

Figures~\ref{fig:SISGaussian} and~\ref{fig:SIGaussian} present results obtained for the above-mentioned aspects when different values of $I_{i,0}$, $\beta_i$ and $\gamma_i$ are used for SIS and SI models respectively. The $\beta_i$ and $\gamma_i$ values are assumed to not vary across rooms. Fig.~\ref{fig:SISGaussian} is obtained when $\gamma=0.75$ while Fig.~\ref{fig:SIGaussian} is obtained when $\gamma=0.0$. From the Fig.~\ref{fig:SISGaussian}, as per SIS model, when there is no infection, dissemination of infection does not occurs because subsequent $dI_{i,t}/dt=0$ (see fig.~\ref{fig:SISGaussian}(a)). This lead to ratio of median risk scores (represented as $r_{m,Infected}/r_{m,Susceptible}$) to be 0 as there are no infected people (see fig.~\ref{fig:SISGaussian}(b)). The inset fig.~\ref{fig:SISGaussian}(b') shows the median risk scores of susceptible people ($r_{m,Susceptible}$) and indicates that, even when there are no confirmed infections, crowded situations which carry high risk can be identified as having high risk scores. 
As our risk model is not dependent on $\beta_i$, $\gamma_i$, and initial infection, via risk score, we are able to detects potential risky situations (see fig.~\ref{fig:SISGaussian}(c),~\ref{fig:SISGaussian}(f), and~\ref{fig:SISGaussian}(i)) which epidemic models such as SIS model are not able to detect. For cases when $I_{i,0}\geq0.01$ and $\beta_i\in\{0.0, 0.5, 1.0\}$, we observe that infections either die off (for $\beta_i=0.0$) or achieve stability (for $\beta_i\in\{0.5, 1.0\}$, see fig.~\ref{fig:SISGaussian}(d) and~\ref{fig:SISGaussian}(g)). The reason for reduction in infections is the recovery rate, while for stability it is the low number of people present when $epoch>1500$. The ratio of median risk scores for different $\beta_i$ and $I_{i,0}\geq0.01$ is shown in fig.~\ref{fig:SISGaussian}(e) and~\ref{fig:SISGaussian}(h)) where we observe that after few epochs the ratio is $<2$ and even reaches $<1$ in short duration. This behavior is because (a) the median value of infected identified by SIS model is less than the median value that of susceptible people and (b) the number of infected is less that number of susceptible. From the fig.~\ref{fig:SISGaussian}(f) and~\ref{fig:SISGaussian}(g) we also see that, irrespective of the epidemic state, most of the people are at high risk.

On the other hand, from  Fig.~\ref{fig:SIGaussian}, we see that when there is no recovery (i.e., $\gamma_i=0.0$) and when $I_{i,0}=0$, the behavior is similar to previous scenario (see fig.~\ref{fig:SIGaussian}(a),~\ref{fig:SIGaussian}(b),~\ref{fig:SIGaussian}(b'),~\ref{fig:SIGaussian}(c),~\ref{fig:SIGaussian}(f), and~\ref{fig:SIGaussian}(i)). Nonetheless, when $I_{i,0}>0.0$ and $\beta_i\neq0.0$, the infections eventually reach entire population which is true as there is no recovery (see fig.~\ref{fig:SIGaussian}(d) and~\ref{fig:SIGaussian}(g)). Further, in this case, due to the above-mentioned reason, ratio of median risk scores is also high (see fig.~\ref{fig:SIGaussian}(e) and~\ref{fig:SIGaussian}(h)). Ratio equal to 1 is achieved when $\beta_i=0$.

\section{Risk score implementation using smartphones}~\label{sec:middleware}
In this section, we demonstrate how the proposed risk score is implemented as part of a smartphone based infection tracking applications. There are two main components required for estimation of the risk score on a smartphone - \textit{(a)} aggregation of risk scores of neighboring smartphones, and \textit{(b)} computation of the risk score of the smartphone itself. Similar to the infection tracking applications used for COVID-19, we consider that estimation of the exposure is done using BLE. However, unlike existing applications which use centralized data repositories to obtain risk scores of neighboring smartphones (i.e. if they are confirmed to be infected), using our approach, each smartphone \textit{(a)} periodically broadcasts its own risk score value, by embedding this value in the BLE advertising packets, and \textit{(b)} periodically updates its own risk score by aggregating the risk scores of all other smartphones in its neighborhood. Such an approach has the following advantages:
\begin{itemize}
    \item The risk score computation does not need to depend on a centralized database containing information about infected individuals, which might be outdated.
    \item The risk score reflect encounters not just with confirmed individuals, but also present environments that are risky from the perspective of infection spread.
    \item Our approach is better suited for privacy preservation, since no information pertaining to the identity of individuals is stored or communicated. 
\end{itemize}

Next, we provide details on the design of the BLE advertising packet as well as how risk score computation is done individually by the smartphone application. For the purpose of this discussion, we refer to the smartphone performing the risk computation as the ego node, and all other smartphones in its vicinity as neighbor nodes. 

\begin{figure}
    \includegraphics[width=\columnwidth]{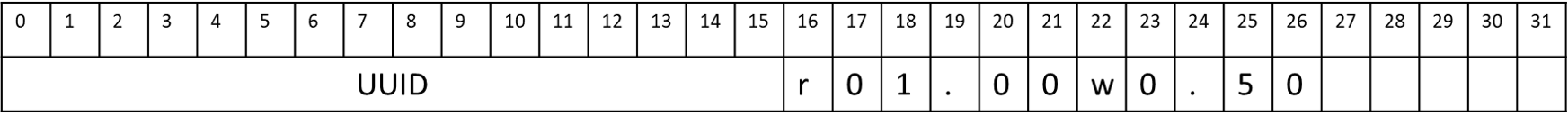}
    \caption{BLE Packet Format}
    \label{fig:BLEPkt}
\end{figure}

\subsection{BLE advertising packet}\label{sec:BLEpkt}
Existing infection tracking techniques record the BLE Media Access Control (MAC) addresses of nearby smartphones~\cite{kindt2020reliable} and compare them with a centralized database of infected individuals. Instead, we discuss how we use the BLE advertising packet to communicate risk score values.

The BLE advertising packet allows including optional payload of up to $31$ bytes \cite{Bluetoot79}. We use these available bytes for broadcasting the risk score. 
Our payload includes:
\begin{enumerate}
    \item A $128$-bit unique identifier (\textbf{UUID}) which is a fixed value used to identify the service, enabling each smartphone to filter out all nearby beacons broadcasting the risk score.
    \item A $6$ bytes long \textbf{Risk score} which includes the risk score value rounded to two decimal places and prefixed by ``\textbf{r}''.
    \item A $5$ bytes long \textbf{weight of the neighbor} which includes the neighbor weight value rounded to two decimal places and prefixed by ``\textbf{w}''.
\end{enumerate}

Fig.~\ref{fig:BLEPkt} shows the payload format of the BLE advertising packet. 

\subsection{Risk score computation}\label{sec:risk-computation}
In addition to the risk scores obtained from the BLE advertisements from the neighbors, the weight of the neighbors, and exposure caused by the neighbors are also required for the purpose of risk score computation, along with the vulnerability of the node itself. Note that here ``node'' means the smartphone.

\subsubsection{Neighbor exposure}\label{sec:nbrexp-impl}

\begin{table}[]
\begin{tabular}{|l|l|}
\hline
\textbf{RSSI}               & \textbf{Neighbor weight} \\ \hline
$>$ -55 dbm                 & 0.8           \\ \hline
$>$ -63 \& $\leq$ -55 dbm   & 0.5           \\ \hline
$>$ -75 \& $\leq$ -63 dbm   & 0.1           \\ \hline
$\leq$ -75 dbm              & 0.0           \\ \hline
\end{tabular}
\caption{Neighbor exposure estimation from RSSI values of BLE advertisements received from neighbors}
\label{tab:attenuation-nbr}
\end{table}

The exposure from a neighbor is an estimation of the likelihood of a transmission event from a neighbor. For infectious diseases such as COVID-19, the likelihood of transmission increases with close contacts. While BLE signal characteristics, such as received signal strength indication (RSSI) and attenuation, can be used for distance estimation, they are known be noisy estimators~\cite{mackey2020improving}. Hence, for the purpose of estimation of exposure from the neighbor, we use a coarse grained mapping, similar to those used in the Exposure API~\cite{Defineme26}. Based on the existing studies, table~\ref{tab:attenuation-nbr} shows how the exposure values can be mapped from the RSSI values~\cite{leith2020coronavirus}. A higher RSSI values maps to a higher  exposure from a neighbor. 




\subsubsection{Neighbor weight}\label{sec:nbrwt-impl}
The neighbor weight is an estimate of the likelihood of a neighboring node to be infectious, which can depend on a range of factors, such as the prevalence of preexisting diseases, age, etc. If such information is available, the derived neighbor weight is included in the BLE advertising packet. However, while such information may not always be available at an individual level, approximate measures are often available at a population level, which can be used as fixed values for all smartphones in a geographic region. For instance, neighbor weight may be derived from the basic reproduction number (R$_0$)~\cite{Heffernan2005} value for a particular epidemic for a given geographical region.

\subsubsection{Vulnerability}\label{sec:vul-impl}
The vulnerability of the ego node is an estimate of how quickly an individual can recover when exposed to infection, and as with the neighbour weight, this depends on a range of factors such as preexisting conditions, age, etc, as well as the nature of the disease itself \cite{Coronavi16}. When available, such information is incorporated in the computation of the risk score by the ego node. 

Currently, the only data shared between the smartphones are the neighbor weights and risk score values. These values are computed on individual smartphones and shared with the neighbors. As no other parameter is shared other than computed values of neighbor weights and risk score and no other information about the neighbor is shared, we enable privacy by design.  
As a proof of concept implementation, we can also provide an alpha version of a smartphone application upon request for the readers to test.

\section{Future directions}\label{sec:fd}
The proposed risk score can be developed further, both in terms of increasing it's accuracy towards risk estimation, as well as applying it to individual use cases, which we highlight in this section.

\subsection{Increasing accuracy of risk score}
The accuracy of the risk score proposed in this paper can also be increased by incorporating additional contextual information, where available. Some examples of this are:

\begin{itemize}
 
    \item \textbf{Indoor and Outdoor location detection:} The likelihood of infection spread has been known to be higher in indoor environments compared to outdoor~\cite{Smieszek2019}. This can be incorporated into the risk score by first, automatically detecting the indoor/outdoor context~\cite{AgarwalMDM2019}, and secondly, by incorporating it into the risk score itself. 
    
    \item \textbf{Identification of exposure context:} As outlined previously in section~\ref{sec:risk-computation}, by identification of the infection context in real-time, the risk score computation can be made more accurate. This can include detection of respiratory symptoms to better estimate the exposure~\cite{sun2015symdetector,liaqat2018towards}. 

    
\end{itemize}

In addition to the points above, a general challenge with all infection tracking applications is that they do not cater to the entire population, since people may not always have access to smartphones and other IoT devices. 


\subsection{Use cases}
The proposed risk score is applicable towards monitoring and managing the spread of infection for population groups, such as over a geographical region, as well as for individuals.


\begin{enumerate}
    \item \textbf{Risk score at different spatial scale:} The proposed risk score, in addition to computing score of an individual can be used to compute the risk score at any spatial scale (i.e., a country, a city, a building, a house, a room). For instance, considering $A_a$ be the area for which risk score has to be computed, such as a district, and let $L_a$ be the group of people in that region at time $t$. The risk score of region $A_a$ at time $t$ is defined as equation~\ref{eq:location}.
    \begin{equation}\label{eq:location}
        r^{A_a}_t= \frac{\sum_{\forall i\in L_a} r_{i,t}}{||L_a||}
    \end{equation}
    Here, $||L_a||$ represents the number of people in $A_a$. Consider a region, $R$ to be comprised of many $A$s, the total population at time $t$, $N^t$ is thus $\sum_{i\in R} ||L_i||$. 
    Some examples include:
    \begin{enumerate}
        \item \textit{\textbf{Monitoring of geographical regions by government authorities:}} As evidenced by the COVID-19 pandemic, the infection spread often starts from small geographical regions, which can grow exponentially if early actions are not taken. Our proposed risk score can be used to obtain an early estimate of the likelihood of infection transmissions within a geographical region. Subsequently, preventative actions, such as increased testing, can be taken, without even resorting to lockdowns that have economic and social impacts.
        
        \item \textit{\textbf{Monitoring of individual buildings:}} An important aspect of managing the spread of infections is to reduce the likelihood of spread in controlled environments, such as office buildings, hotels, hospitals, etc. In such scenarios, risk score can be used to monitor behavior of individuals within such a region, and take quick actions even before any infection is confirmed. Some examples of such use cases are:
        
        \begin{enumerate}
            \item \textit{Hotels:} Guest movements and interactions among guests at hotels can have significant consequences to the infection spread in a pandemic, as has been seen in the case of COVID-19~\cite{Hotelqua56}. The risk score can be used to act quickly by enforcing close monitoring of the individuals who are found to be in risky situations.
            \item \textit{Hospitals:} In order to handle increasing case loads during a pandemic, hospitals typically have dedicated wards. In such cases, it is critical to minimize the likelihood of transmission from such dedicated wards to other wards in the hospital~\cite{oraby2020analysis}, which can be done through monitoring of the risk scores of patients, doctors and other hospital staff.
            \item \textit{Office buildings:} Managing the recovery from a pandemic is equally important to managing it's spread, and the risk score can be used as a part of the plans used for businesses and office buildings~\cite{COVIDSaf33}.
        \end{enumerate}
        Similar scenarios may be envisioned for other closed environments, such as residential buildings, supermarkets, shopping malls, airports, etc.
    \end{enumerate}
     
    \item \textbf{Individual monitoring:} Risk score can also be used to provide real-time alerts to individuals to take action. For instance, it can be used to provide prompts to wear mask if one is detected to move from a less risky region to more risky one. Further, risk score can be used to provide personalized alerts for individuals. For instance, vulnerable people (i.e. who are likely to be affected more due to pre-existing conditions), can be alerted early by using a lower alert threshold.
\end{enumerate}

\section{Conclusion}\label{sec:conc}


In this paper, we introduced a risk score that estimates infection propagation by leveraging the neighborhood of an individuals at a given time. On top, our risk score also takes into consideration factors transmission likelihood and vulnerability to a disease. Our results show that our risk score is able to capture potential risky situations. To further leverage our risk score we demonstrate how our risk score can be implemented in a contact tracing applications and as a proof of concept make it available upon request. Nonetheless, as future directions, we provide use cases and potential parameters that can be included in the risk score to make it more robust.

.

\section*{Acknowledgements}
The authors contributed equally in the research.

\bibliographystyle{ACM-Reference-Format}
\bibliography{biblio}

\end{document}